\documentclass[final,preprint,aps,showpacs,tightenlines,fleqn]{revtex4}
\usepackage{graphicx}
\usepackage{bm}
\usepackage{dcolumn}

\begin{document}

\bibliographystyle{prsty}


\preprint{PNU-NTG-03/2003}
\title{Effective chiral lagrangian in the chiral limit \\ 
from the instanton vacuum}

\author{Hyun-Ah Choi\footnote{E-mail address: hachoi@pusan.ac.kr}
and Hyun-Chul Kim\footnote{E-mail address:hchkim@pusan.ac.kr}}
\affiliation{Department of Physics, 
Pusan National University,\\
609-735 Busan, Republic of Korea}
\date{August 2003}

\begin{abstract}
We study the effective chiral Lagrangian in the chiral limit from the 
instanton vacuum.  Starting from the nonlocal effective chiral action, 
we derive the effective chiral Lagrangian, using the derivative
expansion to order $O(p^4)$ in the chiral limit.  The low energy constants,
$L_1$, $L_2$, and $L_3$ are determined and compared with various
models and the corresponding empirical data.  The results are in a
good agreement with the data.  We also discuss about the upper limit
of the sigma meson, based on the present results.
\end{abstract}

\pacs{PACS: 12.40.-y, 14.20.Dh\\
Key words: Instanton vacuum, nonlocal chiral quark model, effective
chiral lagrangian, low energy constants}
\maketitle

\section{Introduction}
Chiral perturbation theory ($\chi$PT) was introduced as an effective
field theory of QCD in a very low energy 
regime~\cite{Weinberg:1978kz}.  Based on the chirally invariant
Lagrangian with a set of coefficients, $\chi$PT has been of great
success in describing very low-energy phenomena of the  
strong interaction~\cite{Gasser:1983yg,Gasser:1985}~\footnote{For the
reviews, see for example, Refs.~\cite{Pich:1995bw,Ecker} and
references therein.}.  While the structure of the Lagrangian is
determined by the symmetry pattern in QCD, the coefficients are
unknown.  These unknown coeffeicients, known as the low-energy
constants (LECs), contain microscopic information about the
quark-gluon dynamics which would be in principle determined by QCD.
However, it requires a formidable work to derive them from
QCD and thus is impractical.  In fact, they are fitted to the
experimental data such as $\pi\pi$ scattering and $K_{l4}$
decay~\cite{Gasser:1983yg,Amoros:2000mc} and use them for describing
or predicting other processes.  However, when one goes beyond the
leading order, the number of the LECs start to increase very rapidly.
Hence, it is not feasible to fix all LECs to empirical data.  

There has been a great amount of works on the LECs within various
chiral
models~\cite{Diakonov:tw,Andrianov:ay,Balog:ps,Zuk:ht,Ebert:1985kz,    
Chan:1986jq,Simic:jx,Espriu:1989ff,Holdom:iq,Polyakov:1995vh,Fearing:1994ga,  
Bijnens:1995ww,RuizArriola:gc,Peris:1998nj,Wang:1999cp,Heitger:2000ay,Bolokhov:am, 
Alfaro:2002ny}.  Although 
dynamical ingredients of each model are different, almost all models
are in good agreement with empirical data.  Apart from some
models~\cite{Holdom:iq,Wang:1999cp}, many models are based on local
interactions of quarks and mesons.  While the nonlocality of the quark
can be neglected in the range of quark momenta, for example, $k\ll 
1/\bar{\rho}\simeq 600\,{\rm MeV}$ in which $\bar{\rho}$ denotes the
average size of the instanton, recent works 
on the pion wave functions~\cite{Petrov:1998kg,Praszalowicz:2001wy}
and skewed parton distribution~\cite{Petrov:1998kf} show that it is of
great importance to consider the momentum-dependent quark mass in
order to produce the correct end-point behavior of the quark
virtuality.  Similarly, a very recent study on the effective weak chiral
Lagrangian to order $O(p^2)$ from the instanton
vacuum~\cite{Franz:1999yz,Franz:1999ik} asserts that the nonlocality
of the quark plays an essential role in improving previous
results~\cite{Franz:1999wr} concerning the $\Delta T=1/2$ rule in the
LECs.  Furthermore, an appreciable merit of using the
momentum-dependent quark mass as a regulator was already pointed out
by Ball and Ripka~\cite{BallRipka}.  The momentum-dependent quark mass
provides a consistent regularization of the effective action in which
its real and imaginary parts are treated on the same footing and thus
pertinent observables such as anomalous decays $\pi^0\rightarrow
2\gamma$ are safely recovered even if $M(k)$ acts as a regulator.   

In the present work, we shall investigate the effective chiral
Lagrangian from the instanton vacuum~(see a recent
review~\cite{Diakonov:2002fq}).  We first consider the chiral limit as
well as the absence of the external fields.  In order to take into
account the effect of SU(3)-symmetry breaking, one has to modify the
effective chiral action originally obtained by Diakonov and
Petrov~\cite{Musakhanov:1998wp}.  Moreover, the vector and axial-vector
currents are not conserved in the presence of the nonlocal
interaction.  Thus, we first shall concentrate on the LECs in the chiral
limit.  

The outline of the present paper is as follows: In Section II we
briefly explain the instanton-induced chiral quark model, emphasizing in
particular the momentum dependence of the constituent quark mass
and explain how to perform the derivative expansion in the
presence of the momentum-dependent constituent quark mass.
In section III, we show how to derive the ${\cal O} (p^4)$ effective
chiral Lagrangian, using the derivative expansion.  In section IV, we
dicuss the results.  In section V we draw conclusion and make summary.

\section{Chiral quark model from the instanton vacuum}
The instanton vacuum elucidates one of the most important
low-energy properties of QCD, {\em i.e.} the mechanism of spontaneous
breaking of chiral 
symmetry~\cite{DiakonovPetrov,Diakonov:1995ea,Diakonov:1997sj}  
The Banks-Casher relation~\cite{Banks:1979yr} tells us
that the spectral density $\nu(\lambda)$ of the Dirac operator
at zero modes is proportional to the chiral condensate known as an
order parameter of spontaneous breaking of chiral symmetry:
\begin{equation}
\langle \bar{\psi} \psi\rangle \;=\; -\frac{\pi \nu(0)}{V^{(4)}}.
\end{equation}
The picture of the instanton vacuum provides a good realization of
spontaneous breaking of chiral symmetry. A finite density
of instantons and antiinstantons produces the nonvanishing value of
$\nu(0)$, which triggers the mechanism of chiral symmetry breaking.
The Euclidean quark propagator in the
instanton vacuum acquires the following form with a momentum-dependent
quark mass generated dynamically, identified with the coupling strength
between quarks and Goldstone bosons:
\begin{equation}
S_F (k) \;=\; \frac{\rlap{/}{k}+iM(k)}{k^2 + M^2(k)}
\label{Eq:prop}
\end{equation}
with
\begin{equation}
M(k) \;=\; {\rm const}\cdot \sqrt{\frac{N\pi^2 \bar{\rho}^2}{VN_c}}
F^2 (k\bar{\rho}) = M_0 F^2 (k\bar{\rho}).
\label{Eq:mex}
\end{equation}
The ratio $N/V$ denotes the instanton density at equilibrium and
the $\bar{\rho}$ is the average size of the instanton.  The form
factor function $F(k\bar{\rho})$ is related to the Fourier transform
of the would-be zero fermion mode of individual instantons.  The
instanton density $N/V$ is expressed as a gap equation:
\begin{equation}
\frac{N}{V}=4N_c\int\frac{d^4 k}{(4\pi)^4} \frac{M^2(k)}{k^2+M^2(k)} =
1\,{\rm fm}^{-4}.  
\end{equation}
Taking the average instanton size $\bar{\rho}=1/3$ fm, one obtains
$M_0\simeq 350$ MeV.  

The instanton vacuum induces effective $2N_f$-fermion
interactions~\cite{DiakonovPetrov,Diakonov:1997sj,Diakonov:1995ea}.  
For example, it has a type of the 
Nambu-Jona-Lasinio model for $N_f=2$ while for $N_f=3$ it exhibits the
't~Hooft determinant~\cite{tHooft:1976}.  Goldstone bosons appear as
collective excitations by quark loops generating a dynamic quark mass.
Eventually it is found that at low energies QCD is reduced to an
interacting quark-Goldstone boson theory given by the following
Euclidean partition function\cite{Diakonov:1997sj}
\begin{eqnarray}
{\cal Z}&=& \int {\cal D} \psi {\cal D} \psi^\dagger
{\cal D} \pi^a
\exp \int d^4 x \Big[ \psi^{\dagger \alpha}_{f} (x)
i \rlap{/}{\partial}\psi^{\alpha}_{f}(x)  \cr
&+& i \int \frac{d^4 k d^4 l}{(2\pi)^8} e^{i(k-l)\cdot x}
\sqrt{M(k) M(l)} \psi^{\dagger \alpha}_{f} (k)
\left(U^{\gamma_5}\right)_{fg}
\psi^{\alpha}_{g} (l)\Big],
\label{Eq:Dirac}
\end{eqnarray}
where $U^{\gamma_5}$ stands for the pseudo-Goldstone boson:
\begin{equation}
U^{\gamma_5} (x) = U(x) \frac{1+\gamma_5}{2} + U^\dagger (x)
\frac{1-\gamma_5}{2} \;=\;
\exp\left(i\pi^a(x) \lambda^a\gamma_5 /f_\pi \right).
\end{equation}
The $\alpha$ is the color index, $\alpha = 1,\cdots, N_c$ and
$f$ and $g$ are flavor indices.  $M(k)$ is the constituent quark mass
being now momentum-dependent, which is expressed by Eq.(\ref{Eq:mex}). 
Its momentum dependence will play a main role in the present work. If
we choose $F(k\bar{\rho})$ to be constant and add a regularization 
({\em e.g.} Pauli-Villars or proper-time), the partition function
becomes just that of the usual $\chi$QM.  The original expression for
the $F(k\bar{\rho})$~\cite{DiakonovPetrov}, which is obtained from the
Fourier transformation of the would-be zero fermion mode of individual
instantons with the sharp instanton distribution assumed, is as
follows: 
\begin{equation}
F(k\bar{\rho}) \;=\; 2z \left(
I_0 (z) K_1 (z)
- I_1 (z) K_0 (z)
-\frac{1}{z}
I_1 (z) K_1 (z)\right).
\label{Eq:fofk}
\end{equation}
Here $I_0$, $I_1$, $K_0$, and $K_1$ denote
the modified Bessel functions, z is defined as $z=k\bar{\rho}/2$.  When
$k$ goes to infinity, the form factor $F(k\bar{\rho})$ has the
following asymptotic behavior:
\begin{equation}
F(k\bar{\rho}) \longrightarrow \frac{6}{(k\bar{\rho})^3}.
\end{equation}

Actually, there are other ways of understanding the nonlocal effective
interaction without relying on the instanton
vacuum~\cite{Bowler:ir,Ripka,Alkofer:2000wg,Roberts:2000aa}.  In those
cases, the momentum-dependent quark mass can be interpreted as a
nonlocal regularization in Euclidean space.  Hence, various types of
the $M(k)$ as a regulator with  the regularization parameter
$\Lambda\sim 1/ \bar{\rho}$ has been used by different authors.  For
example, the dipole-type $M(k)$ is used in the study of the pion wave 
function~\cite{Petrov:1998kg}, while the Gaussian is employed in
Ref.~\cite{Golli:1998rf}.    

Therefore, we will not confine ourselves to the expression given
in Eq.(\ref{Eq:fofk}) but rather try three different types of the
$M(k)$:
\begin{equation}
M(k) = \left \{
\begin{array}{l} \mbox{Eqs.(\ref{Eq:mex}, \ref{Eq:fofk})} \\
M_0 \left(\frac{4\Lambda^2}{4\Lambda^2 + k^2}\right)^4 \\
 M_0 \exp{\left(-\frac{k^2}{\Lambda^2}\right)}
\end{array} \right. ,
\label{Eq:types}
\end{equation}
where the cut-off parameter $\Lambda$ is taken as the inverse of
$\bar{\rho}$.  The $M(k)$ is normalized to $M_0$ at $k=0$.
Originally, $M_0$ is found to be around $350$ MeV.  However, we will
regard $M_0$ as a free parameter ranging from $200$ MeV  
to $450$ MeV and fit for each $M_0$ the parameter $\Lambda$ to
the pion decay constant $f_\pi=93$ MeV.  Figure~1 shows the
momentum dependence of the three different types of $M(k)$ with
$M_0=350$~MeV.  The dipole type displays the largest tale, while the Gaussian
takes the strongly suppressed tail, compared to other ones.  As will be shown
later, this difference appearing in the tail is basically
responsible for the different results in the LECs of the effective weak
chiral Lagrangian.
\begin{figure}
\begin{center}
\includegraphics[width=10cm,height=8.0cm]{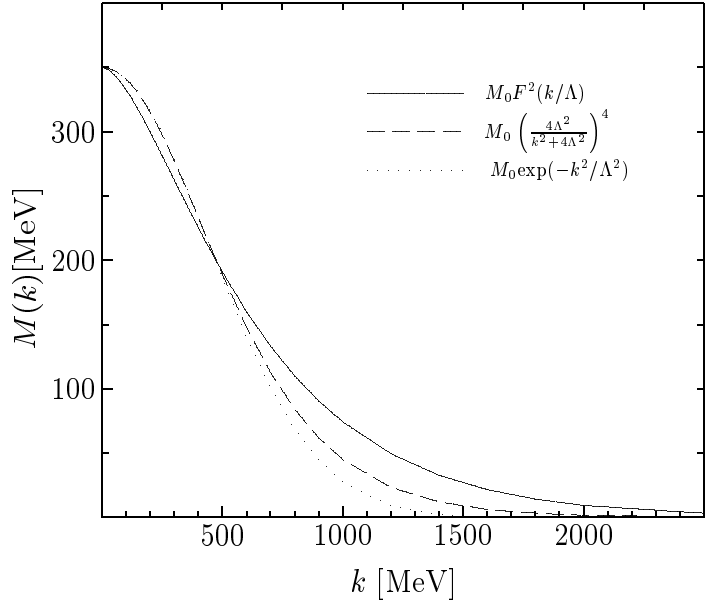}
\caption{The dependence of the $M(k)$ on $|k|$.  The solid curve draws
the Diakonov-Petrov $M(k)$, the dashed one shows the dipole-type
parametrization of $M(k)$, and the dotted one corresponds to the
Gaussian type of $M(k)$.}
\end{center}
\end{figure}

This effective theory of quarks and light Goldstone mesons applies
to quark momenta up to the inverse size of the instanton,
$\bar{\rho}^{-1} \simeq 600$ MeV, which may act as a
scale of the model ($\mu_{\chi{\rm QM}}$).  A merit to derive the
$\chi$QM from the instanton vacuum lies in the fact that the scale of the
model is naturally determined by $\bar{\rho}^{-1}$.  Furthermore,
mesons and baryons can be treated on the same footing
in the $\chi$QM.  For example, the model has been very successful
in describing the properties of the baryons~\cite{christov:1995}.

\section{Effective chiral Lagrangian to order ${\cal O}(p^4)$}
The low-energy effective QCD partition function given in
Eq.~(\ref{Eq:Dirac}) is the starting point of the present work.
Having integrated out the quark fields of Eq.~(\ref{Eq:Dirac}),  
we obtain
\begin{equation}
{\cal Z} \;=\; \int {\cal D} \pi^a \exp{\left(-S_{\rm eff}[\pi^a]\right)},
\end{equation}
where the $S_{\rm eff}[\pi^a]$ stands for the effective chiral action:
\begin{equation}
S_{\rm eff} [\pi^a] = -N_c \ln {\rm det} D (U^{\gamma_5}).
\label{Eq:det}
\end{equation}
Here, the $D(U^{\gamma_5})$ is the Dirac operator defined by
\begin{equation}
D \;=\; i\rlap{/}{\partial} + i\sqrt{M(-i\partial)} U^{\gamma_5}
\sqrt{M(-i\partial)}.
\label{eq:dirac}
\end{equation}
The Dirac operator is not Hermitian, so that it is useful to
divide the effective action into the real and imaginary parts:
\begin{eqnarray}
{\rm Re} S_{\rm eff} &=& \frac12 \left(S_{\rm eff} + S_{\rm eff}^*\right)
\;=\; -\frac12 N_c \ln {\rm det} \left[D^\dagger D\right],
\label{Eq:realech}
\\
i{\rm Im} S_{\rm eff} &=& \frac12 \left(S_{\rm eff} - S_{\rm eff}^*\right)
\;=\; -\frac12 N_c \ln {\rm det} \left[D /D^\dagger\right].
\end{eqnarray}

It is already known that the imaginary part of the effective chiral action
is identical to the Wess-Zumino-Witten action~\cite{Wess:yu,Witten:tx}
with the correct coefficient, which arises from the derivative expansion
of the imaginary part to ${\cal O}(p^5)$~\cite{AF,Chan,DSW,Diakonov:1987ty}.
An appreciable merit of using the momentum-dependent
quark mass as a regulator was already pointed out by
Ball and Ripka~\cite{BallRipka}.  The momentum-dependent quark mass
provides a consistent regularization of the effective action given in
Eq.(\ref{Eq:det}) in which its real and imaginary parts
are treated on the same footing and thus pertinent observables
such as anomalous decays $\pi^0\rightarrow 2\gamma$ are safely recovered
even if $M(k)$ acts as a regulator.  Hence, in this work, we will
concentrate on the real part of the effective chiral action which will
provide us with the effective chiral Lagrangian with the LECs
determined.  In the present work, we first consider the case of the
chiral limit and turn off the external fields.  Furthermore, we keep
only the leading order in the large $N_c$.   

In order to calculate the real part given in Eq.(\ref{Eq:realech}), we substract
the vacuum part and use the derivative expansion. We therefore write
\begin{eqnarray}
&& {\rm Re} S_{\rm eff}[\pi^a] - {\rm Re} S_{\rm eff}[0]\cr
&=& -\frac{N_c}{2} {\rm Tr} \ln
\left(\frac{D^\dagger D}{D^{\dagger}_{0} D_0} \right) \cr
&=&  -\frac{N_c}{2} \int d^4 x \int \frac{d^4 k}{(2\pi)^4}
e^{-i k x} {\rm tr} \ln
\left(\frac{D^\dagger D}{D^{\dagger}_{0} D_0} \right) e^{i k x} \cr
&=& -\frac{N_c}{2} \int d^4 x \int \frac{d^4 k}{(2\pi)^4}
{\rm tr} \ln
\left(\frac{D^\dagger(\partial \rightarrow \partial + i k)
 D(\partial \rightarrow \partial + i k)}
{D^{\dagger}_{0}(\partial \rightarrow \partial + i k)
D_0(\partial \rightarrow \partial + i k)} \right) \cdot 1\cr
&=&-\frac{N_c}{2} \int d^4 x \int \frac{d^4 k}{(2\pi)^4} {\rm tr~ln} 
\left(1-\frac{N}{D_0 ^\dag (\partial + ik) D_0 (\partial + ik)}
\right) \cdot 1 \cr  
&=& \frac{N_c}{2} \int d^4 x \int \frac{d^4 k}{(2\pi)^4} 
{\rm tr~ln}\left(\frac{1}{D_0 ^\dag D_0 } N + \frac{1}{2} 
\frac{1}{D_0 ^\dag D_0} N \frac{1}{D_0 ^\dag  D_0} N \right. \cr \cr
&& + \left. \frac{1}{3}\frac{1}{D_0 ^\dag  D_0} N \frac{1}{D_0
    ^\dag D_0 } 
N \frac{1}{D_0 ^\dag D_0 } N 
+ \frac{1}{4} \frac{1}{D_0 ^\dag  D_0 } N \frac{1}{D_0 ^\dag  D_0 } N 
\frac{1}{D_0 ^\dag  D_0 } N \frac{1}{D_0 ^\dag  D_0 } N + \dots \right) \cdot 1,
\label{Eq:Seff}
\end{eqnarray}
where 
\begin{equation}
N=D_{0}^{\dag}({\partial}+ik)D_{0}({\partial}+ik)
-D^{\dag}({\partial}+ik)D({\partial}+ik) .  
\end{equation}
Here we have used a complete set of plane waves for the calculation of the
functional trace, summing over all states and taking the trace in $x$.
'tr' then denotes the usual matrix trace over flavor and Dirac
spaces.  The RHS of Eq.(\ref{Eq:Seff}) can now be expanded 
in powers of the derivatives of the pseudo-Goldstone boson  fields,
$\rlap{/}{\partial} U^{\gamma_5}$, and of $2ik \cdot \partial+\partial^2$.
The operators $D^\dagger D$ and $D_0^\dagger D_0$ in
Eq.(\ref{Eq:Seff}) can be expanded as follows: 
\begin{eqnarray}
D^{\dag} ({\partial} +ik) D({\partial} +ik)
&=& -{\partial}^{2} -2ik \cdot {\partial} +k^{2}
-\sqrt{M(-i{\partial}+k)} ({/\hspace{-0.2 cm}\partial}U^{\gamma_5})
\sqrt{M(-i{\partial}+k)} \cr 
&& +\sqrt{M(-i{\partial}+k)} U^{-\gamma_5} M(-i{\partial}+k)
U^{\gamma_5} \sqrt{M(-i{\partial}+k)},\label{eq:dd}\\
D_0 ^{\dag} ({\partial} +ik) D_0 ({\partial} +ik) &=&
  -{\partial}^{2} -2ik \cdot {\partial} +k^{2}
  +M^{2}(-i{\partial}+k).
\label{eq:dd0}
\end{eqnarray}
Since the dynamic quark mass in Eqs.(\ref{eq:dd},\ref{eq:dd0})
contains the derivatives, we need to expand it to order ${\cal O}
(\partial^4)$: 
\begin{eqnarray}
\lefteqn{\sqrt{M(-{\partial}^{2}-2ik \cdot {\partial}+k^{2})} }\cr
&=& \sqrt{M(k^{2})} \left(1-\frac{{\tilde{M}'}}{2M}{\partial}^{2}
  -\frac{(\tilde{M}')^{2}}{8M^{2}}{\partial}^{4} 
+\frac{{\tilde{M}''}}{4M}{\partial}^{4}
  -i\frac{{\tilde{M}'}}{M}k_{\alpha}{\partial}_{\alpha} \right. \cr 
&&
  -i\frac{(\tilde{M}')^{2}}{2M^{2}}k_{\alpha}{\partial}_{\alpha}{\partial}^{2}
  +i\frac{{\tilde{M}''}}{M}k_{\alpha}{\partial}_{\alpha}{\partial}^{2} 
+\frac{(\tilde{M}')^{2}}{2M^{2}}k_{\alpha}k_{\beta}{\partial}_{\alpha}{\partial}_{\beta} 
-\frac{{\tilde{M}''}}{M}k_{\alpha}k_{\beta}{\partial}_{\alpha}{\partial}_{\beta} \cr
&&
  +\frac{3(\tilde{M}')^{3}}{4M^{3}}k_{\alpha}k_{\beta}{\partial}_{\alpha}
{\partial}_{\beta}{\partial}^{2} 
-\frac{3{\tilde{M}'}{\tilde{M}''}}{2M^{2}}k_{\alpha}k_{\beta}{\partial}_{\alpha}
{\partial}_{\beta}{\partial}^{2}+\frac{{\tilde{M}'''}}{M}k_{\alpha}k_{\beta}
{\partial}_{\alpha}{\partial}_{\beta}{\partial}^{2}  \cr &&
  +i\frac{(\tilde{M}')^{3}}{2M^{3}}k_{\alpha}k_{\beta}k_{\rho}
{\partial}_{\alpha}{\partial}_{\beta}{\partial}_{\rho}
  -i\frac{{\tilde{M}'}{\tilde{M}''}}{M^{2}}k_{\alpha}k_{\beta}k_{\rho}
{\partial}_{\alpha}{\partial}_{\beta}{\partial}_{\rho}
  \cr &&
  +i\frac{2{\tilde{M}'''}}{3M}k_{\alpha}k_{\beta}k_{\rho}{\partial}_{\alpha}
{\partial}_{\beta}{\partial}_{\rho}
  -\frac{5(\tilde{M}')^{4}}{8M^{4}}k_{\alpha}k_{\beta}k_{\rho}k_{\sigma}
{\partial}_{\alpha}{\partial}_{\beta}{\partial}_{\rho}{\partial}_{\sigma}
  \cr &&
  +\frac{3(\tilde{M}')^{2}{\tilde{M}''}}{2M^{3}}k_{\alpha}k_{\beta}k_{\rho}
k_{\sigma}{\partial}_{\alpha}{\partial}_{\beta}{\partial}_{\rho}{\partial}_{\sigma}
  -\frac{(\tilde{M}'')^{2}}{2M^{2}}k_{\alpha}k_{\beta}k_{\rho}k_{\sigma}
{\partial}_{\alpha}{\partial}_{\beta}{\partial}_{\rho}{\partial}_{\sigma}\cr  
&&
\left. -\frac{2{\tilde{M}'}{\tilde{M}'''}}{3M^{2}}k_{\alpha}k_{\beta}
k_{\rho}k_{\sigma}{\partial}_{\alpha}{\partial}_{\beta}{\partial}_{\rho}
{\partial}_{\sigma} +\frac{{\tilde{M}''''}}{3M}k_{\alpha}k_{\beta}
k_{\rho}k_{\sigma}{\partial}_{\alpha}{\partial}_{\beta}
{\partial}_{\rho}{\partial}_{\sigma}  \right) \cr 
&& + {\cal O}({\partial}^{5}),\\
\lefteqn{M(-{\partial}^{2}-2ik \cdot {\partial}+k^{2})} \cr 
&=& M(k^{2}) -{\tilde{M}'}{\partial}^{2}
  -2{\tilde{M}''}k_{\alpha}k_{\beta}{\partial}_{\alpha}{\partial}_{\beta}
+\frac{1}{2}{\tilde{M}''}{\partial}^{2}  \cr 
&& +2{\tilde{M}'''}k_{\alpha}k_{\beta}{\partial}_{\alpha}{\partial}_{\beta}{\partial}^{2}
+\frac{2}{3}{\tilde{M}''''} 
k_{\alpha}k_{\beta}k_{\rho}k_{\sigma}{\partial}_{\alpha}
{\partial}_{\beta}{\partial}_{\rho}{\partial}_{\sigma}
  \cr 
&&  -2i{\tilde{M}'}k_{\alpha}{\partial}_{\alpha}+2i{\tilde{M}''}
k_{\alpha}{\partial}_{\alpha}{\partial}^{2} 
+\frac{4}{3}i{\tilde{M}'''}k_{\alpha}k_{\beta}k_{\rho}
{\partial}_{\alpha}{\partial}_{\beta}{\partial}_{\rho}+{\cal O}({\partial}^{5}),\\
\lefteqn{M^{2}(-{\partial}^{2}-2ik \cdot {\partial}+k^{2})} \cr  
&=&   M^{2}(k^{2})-2M{\tilde{M}'}{\partial}^{2}+(\tilde{M}')^{2}
{\partial}^{4}+M{\tilde{M}''}{\partial}^{4}  \cr 
&&
  -4iM{\tilde{M}'}k_{\alpha}{\partial}_{\alpha}+4i(\tilde{M}')^{2}
k_{\alpha}{\partial}_{\alpha}{\partial}^{2}
  +4iM{\tilde{M}''}k_{\alpha}{\partial}_{\alpha}{\partial}^{2} \cr 
&& -4(\tilde{M}')^{2}k_{\alpha}k_{\beta}{\partial}_{\alpha}
{\partial}_{\beta}-4M{\tilde{M}''}k_{\alpha}k_{\beta}
{\partial}_{\alpha}{\partial}_{\beta} 
+12{\tilde{M}'}{\tilde{M}''}k_{\alpha}k_{\beta}{\partial}_{\alpha}
{\partial}_{\beta}{\partial}^{2}  \cr &&
  +4M{\tilde{M}'''}k_{\alpha}k_{\beta}{\partial}_{\alpha}
{\partial}_{\beta}{\partial}^{2}
  +8i{\tilde{M}'}{\tilde{M}''}k_{\alpha}k_{\beta}k_{\rho}
{\partial}_{\alpha}{\partial}_{\beta}{\partial}_{\rho}
  \cr &&
  +\frac{8}{3}iM{\tilde{M}'''}k_{\alpha}k_{\beta}k_{\rho}
{\partial}_{\alpha}{\partial}_{\beta}{\partial}_{\rho}
  +4(\tilde{M}'')^{2}k_{\alpha}k_{\beta}k_{\rho}k_{\sigma}
{\partial}_{\alpha}{\partial}_{\beta}{\partial}_{\rho}{\partial}_{\sigma}
  \cr && 
  +\frac{16}{3}{\tilde{M}'}{\tilde{M}'''}k_{\alpha}k_{\beta}k_{\rho}
k_{\sigma}{\partial}_{\alpha}{\partial}_{\beta}{\partial}_{\rho}{\partial}_{\sigma}
  +\frac{4}{3}M{\tilde{M}''''}k_{\alpha}k_{\beta}k_{\rho}k_{\sigma}
{\partial}_{\alpha}{\partial}_{\beta}{\partial}_{\rho}{\partial}_{\sigma}  \cr  
&& + {\cal O}({\partial}^{5}),
\end{eqnarray}
where
\begin{eqnarray}
\label{Eq:Ms}
&&M=M(k), \;\; \tilde{M}' = \frac{1}{2k}\frac{dM(k)}{dk}=\frac{1}{2k}M^{'}(k), \cr
&& \tilde{M}''=\frac{1}{4k^{3}}\left(\frac{d^{2} M(k)}{dk^{2}}k -
  \frac{d M(k)}{dk}\right)=\frac{1}{4k^{3}}(M^{''}(k)k-M^{'}(k)) , \cr 
&& \tilde{M}'''=\frac{1}{8k^{5}}\left(k^2 \frac{d^{3}M}{dk^{3}}-3k 
\frac{d^{2}M}{dk^{2}}+3\frac{dM}{dk}\right) \cr
&& =\frac{1}{8k^{5}}(M^{'''}(k)k^{2}-3M^{''}(k)k+3M^{'}(k)), \cr
&&\tilde{M}''''=\frac{1}{16k^{7}}\left(k^{3}\frac{d^{4}M}{dk^{4}}
-6k^{2}\frac{d^{3}M}{dk^{3}}+15k
\frac{d^{2}M}{dk^{2}}-15\frac{dM}{dk}\right) , \cr 
&& =\frac{1}{16k^{7}}(M^{''''}(k)k^{3} - 6M^{'''}(k)k^{2}
+ 15M^{''}(k)k - 15M^{'}(k)) .
\end{eqnarray}

Having carried out the necessary arithmetic and grouped terms for each
order in the meson momentum, we finally obtain the effective chiral
Lagrangian to order ${\cal O}(p^4)$ with the momentum-dependent quark
mass.  The effective chiral Lagrangian ${\cal L }^{(2)}$ to order
${\cal O}(p^2)$ is given as follows:  
\begin{equation}
{\cal L }^{(2)} = \frac{f_{\pi} ^{2}}{4} \left \langle \partial_{\mu}
  U^{\dagger} \partial_{\mu} U \right \rangle.
\label{Eq:SeffFpi}
\end{equation}
In Eq.(\ref{Eq:SeffFpi}) $\langle \rangle$ denotes the flavor trace   
and $f_{\pi}$ is the well-known pion decay constant $f_\pi=93$ MeV
expressed by 
\begin{equation}
f_{\pi}^2 \;=\; 4 N_c \int \frac{d^4 k}{(2\pi)^4}
\frac{M^2(k) - \frac12 M(k) M'(k)k
+ \frac14 M'^2(k) k^2}{(k^2 +M^2(k))^2}.
\label{Eq:fpiq}
\end{equation}
Equation~(\ref{Eq:fpiq}) has been already derived (see, for example,
Refs.~\cite{Bowler:ir,Golli:1998rf}).  We will use Eq.(\ref{Eq:fpiq})
to fix the cut-off parameter $\Lambda$.  When we switch off the
momentum dependence of the constituent quark mass,  
we end up with the well-known expression of the $\chi$QM for $f^2_{\pi}$:
\begin{equation}
f_{\pi}^2 \;=\; 4 N_c^{2}\int \frac{d^4 k}{(2\pi)^4}
\frac{M^2}{(k^2 +M^2)^2}, \; M=\mbox{ const},
\end{equation}
which is logarithmically divergent.  

The ${\cal O} (p^4)$ effective chiral Lagrangian in the chiral limit
is obtained as follows: 
\begin{equation}
{\cal L }^{(4)}=L_1 \left \langle \partial_{\mu} U^{\dagger}
  \partial_{\mu}U\right\rangle^2  + L_2 \left \langle \partial_{\mu}
U^{\dagger} \partial_{\nu} U \right \rangle^2  
+ L_3 \left \langle \partial_{\mu} U^{\dagger} \partial_{\mu} U
  \partial_{\nu} U^{\dagger} \partial_{\nu} U \right \rangle . 
\end{equation}
where $L_1$, $L_2$, and $L_3$ denote the LECs for the ${\cal O} (p^4)$
effective chiral Lagrangian:
\begin{eqnarray}
L_1 &=& \frac{N_c}{4} \int \frac{d^4 k}{(2\pi)^4}
  \frac{1}{(k^{2}+M^{2})^{4}} \left[ M^{4} + \frac{1}{6} M^{4}M^{'2} +
  \frac{1}{24} M^{4}M^{'4} \right. \cr  &&
  - \frac{1}{6} M^{5}M^{''} - \frac{1}{24} M^{5}M^{'2}M^{''} 
- \frac{1}{2k} M^{5}M^{'} -
 \frac{1}{24k} M^{5}M^{'3} \cr  &&
 - \frac{1}{2} kM^{3}M^{'} + \frac{7}{12}kM^{3}M^{'3} 
- \frac{1}{6} kM^{4}M^{'}M^{''} - \frac{1}{4} k^{2}M^{2}M^{'4} \cr  &&
-\frac{1}{6} k^{2}M^{3}M^{''} + \frac{1}{12} k^{2}M^{3}M^{'2}M^{''}  -
\frac{1}{24} k^{3}M M^{'3} - \frac{1}{6} k^{3}M^{2}M^{'}M^{''}  \cr &&
\left. + \frac{1}{8} k^{4}M M^{'2}M^{''} \right],\label{eq:lecs1}\\  
L_2 &=& 2 L_1 ,\label{eq:lecs2} \\  
L_3 &=& N_c \int \frac{d^4 k}{(2\pi)^4} \frac{1}{(k^{2}+M^{2})^{4}} \left[
  - M^{4} - \frac{13}{16} M^{4}M^{'2} - \frac{1}{8} M^{4}M^{'4} \right. \cr  &&
  + \frac{53}{96} M^{5}M^{''} + \frac{3}{16} M^{5}M^{'2}M^{''} + 
\frac{41}{32k} M^{5}M^{'} +
  \frac{3}{16k} M^{5}M^{'3} \cr  &&
- \frac{19}{32} kM^{3}M^{''} - \frac{3}{4}kM^{3}M^{'3}  
- \frac{1}{8} kM^{4}M^{'}M^{''} + \frac{3}{8} k^{2}M^{2}M^{'4} \cr  &&
  + \frac{41}{96} k^{2}M^{3}M^{''} + \frac{1}{16} k^{3}MM^{'3} 
+ \frac{1}{16} k^{3}M^{2}M^{'}M^{''} - \frac{3}{16} k^{4}MM^{'2}M^{''} \cr  &&
  - \frac{1}{32} M^{6}M^{''2} - \frac{1}{24} M^{6}M^{'}M^{'''}
 + \frac{1}{96} M^{7}M^{''''} - \frac{1}{32k^{3}} M^{7}M^{'} \cr 
&& - \frac{1}{32k^{2}} M^{6}M^{'2} + \frac{1}{32k^{2}} M^{7}M^{''} 
- \frac{3}{16k} M^{6}M^{'}M^{''} + \frac{1}{16k} M^{7}M^{'''} \cr  &&
+ \frac{3}{16} kM^{5}M^{'''} +  \frac{23}{32} k^{2}M^{2}M^{'2} 
- \frac{1}{16} k^{2}M^{4}M^{''2} - \frac{1}{12}k^{2}M^{4}M^{'}M^{'''} \cr  &&
+ \frac{1}{32} k^{2}M^{5}M^{''''} + \frac{3}{32} k^{3}M M^{'} 
+ \frac{3}{16} k^{3}M^{3}M^{'''} - \frac{3}{32} k^{4}M M^{''} \cr  &&
- \frac{1}{32} k^{4}M^{2}M^{''2} - \frac{1}{24} k^{4}M^{2}M^{'}M^{'''} 
+ \frac{1}{32} k^{4}M^{3}M^{''''} + \frac{1}{16} k^{5}M M^{'''} \cr  &&
\left.+ \frac{1}{96} k^{6}M M^{''''} \right]. 
\label{eq:lecs3}
\end{eqnarray}
Equation~(\ref{eq:lecs2}) is the large-$N_c$ relation which was
derived from the OZI rule for the meson scattering
amplitude~\cite{Gasser:1985}.  If we turn off the momentum-dependence
of the constituent quark mass, we reproduce the results of the
usual $\chi$QM.  Eqs.(\ref{eq:lecs1}-\ref{eq:lecs3}) are our main
results.    
\section{Results and discussion}
The parameters in the present model are the constituent quark mass
$M_0$ at $k^2=0$ and the cut-off parameter $\Lambda$ in
Eq.(\ref{Eq:types}).  The cut-off parameter $\Lambda$ is fixed 
by reproducing the pion decay constant $f_\pi^{2}$.  
Having chosen the $\Lambda$, we are able to calculate the LECs 
$L_1$, $L_2$, and $L_3$, numerically.  The only free parameter we have
is the $M_0$.  In Table I, the results of the $L_1$, $L_2$, and $L_3$
are listed with $M_0=350$ MeV.  The results are found to be rather  
insensitive to the types of $M(k)$.  They are compared with those from other models.
In Table I, GL denotes the empirical data obtained by 
Gasser and Leutwyler~\cite{Gasser:1983yg}.  The results are in a good
agreement with Ref.~\cite{Bijnens:1995ww,Gasser:1983yg}.  It is
interesting to compare the present results with those from 
Ref.~\cite{Holdom:iq}, since it emphasizes also the
momentum-dependence of the quark mass.  Holdom {\em et
  al.}~\cite{Holdom:iq} used two different values of the quark
self-energy $\Sigma(p)$.  Holdom (1) represents the quark self-energy
$\Sigma(p)_1=\frac{2M^3}{M^2 + p^2}$, while Holdom (2) designates
$\Sigma(p)_2=\frac{4M^3}{3M^2 + p^2}$.  $M$ denotes the constituent quark mass.  
\begin{table}[d]
\caption{The low energy constants $L_1$, $L_2$, $L_3$.}
\begin{tabular}{cccccc} \\ \hline \hline
& $M_0$(MeV) & $\Lambda $(MeV) &$L_1 ( \times 10^{-3})$&
$L_2 ( \times 10^{-3})$ &$L_3 ( \times 10^{-3})$ \\ \hline 
local $\chi$QM & $350$ & $1905.5$ & $0.79$ & $1.58$ & $-3.17$ \\ \hline
DP  & $350$ & $611.7$& $0.82$ & $1.63$ & $-3.09$ \\ 
Dipole & $350$ & $611.2$ & $0.82$ & $1.63$ & $-2.97$ \\  
Gaussian & $350$ & $627.4$ & $0.81$ & $1.62$ & $-2.88$ \\ \hline
GL & & & $0.9\pm 0.3$ & $1.7\pm 0.7$ & $-4.4 \pm 2.5$ \\
Bijnens &  & & $0.6 \pm 0.2$ & $1.2 \pm 0.4$ & $-3.6 \pm 1.3$ \\ 
Arriola &  & & $0.96$ & $1.95$ & $-5.21$  \\
VMD & & & $1.1$ & $2.2$ & $-5.5$ \\
Holdom(1) &  & & $0.97$ & $1.95$ & $-4.20$ \\
Holdom(2) & & & $0.90$ & $1.80$ & $-3.90$ \\ 
Bolokhov et al. & & & $0.63$ & $1.25$ & $2.50$ \\
Alfaro et al. & & & $0.45$ & $0.9$ & $-1.8$ \\ 
\hline
\end{tabular} 
\end{table} 

\begin{figure}
\begin{center}
\includegraphics[height=7cm]{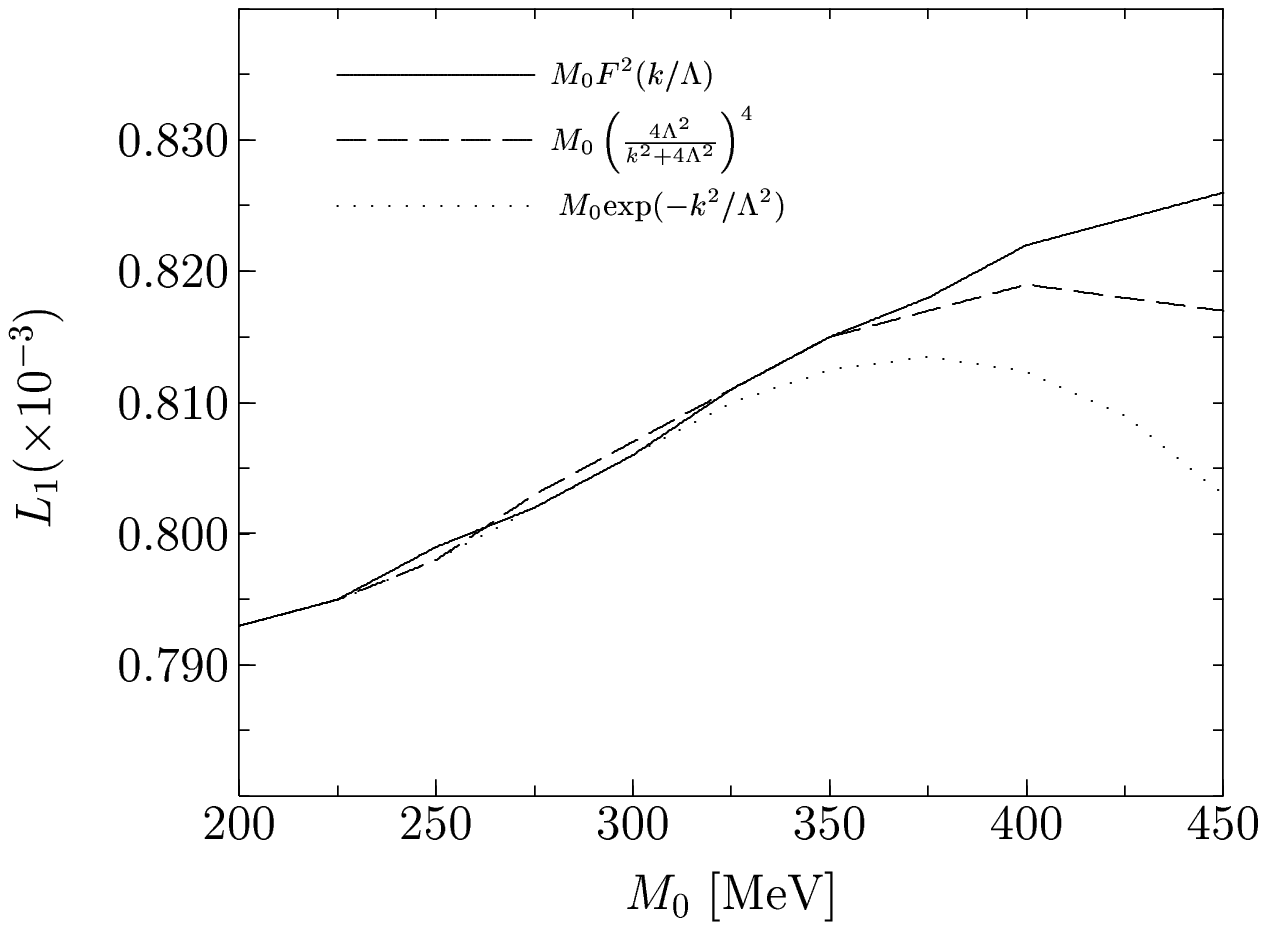}
\caption{The dependence of $L_1$ on $M_0$.  The solid curve stands for
the result with the form factor from the instanton vacuum
given in Eq.(\ref{Eq:types}),
the dashed one draws the result with the dipole type $M(k)$,
and the dotted one designates the result with the Gaussian one.}
\end{center}
\end{figure}

\begin{figure}
\begin{center}
\includegraphics[height=7cm]{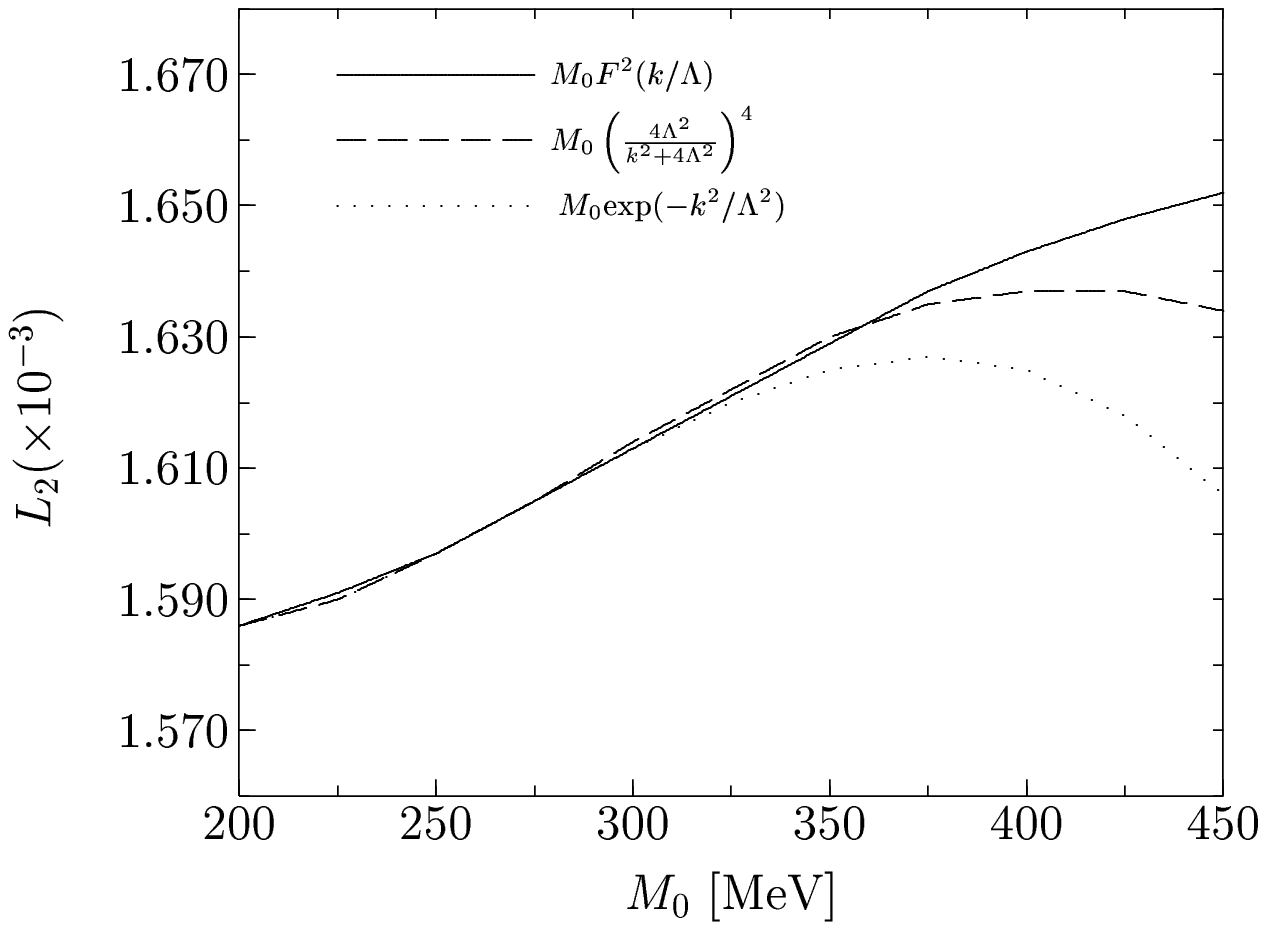}
\caption{The dependence of $L_2$ on $M_0$.  The solid curve stands for
the result with the form factor from the instanton vacuum
given in Eq.(\ref{Eq:types}),
the dashed one draws the result with the dipole type $M(k)$,
and the dotted one designates the result with the Gaussian one.}
\end{center}
\end{figure}

\begin{figure}
\begin{center}
\includegraphics[height=7cm]{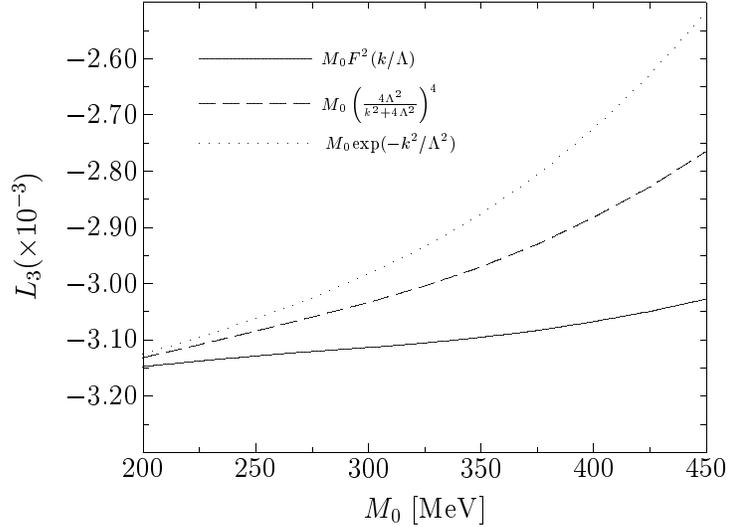}
\caption{The dependence of $L_3$ on $M_0$.  The solid curve stands for
the result with the form factor from the instanton vacuum
given in Eq.(\ref{Eq:types}),
the dashed one draws the result with the dipole type $M(k)$,
and the dotted one designates the result with the Gaussian one.}
\end{center}
\end{figure}

Figures 2, 3 and 4 draw the dependence of the $L_1$, $L_2$, and $L_3$ on the
$M_0$, respectively.  While the results with three different $M(k)$ 
show a similar behavior in smaller $M_0$, they become rather different 
as $M_0$ increases.  In particular, the Gaussian type of $F(k)$
drastically suppresses the LECs at higher values of $M_0$.  The reason
can be found in the behavior of the $M(k)$.  The Gaussian type of
$F(k)$ decreases rather strongly as $k$ increases, compared to other
two different types of form factors.  

Apart from the relation of the large $N_c$ limit, there is an
additional relation in the local $\chi$QM: $2L_2 + L_3=0$.  The
dual-resonance model has the same relation~\cite{Bolokhov:am,Alfaro:2002ny}.
However, the quantity $2L_2+L_3$ is not equal to zero in the present
model.  Interestingly, this relation is deeply related to the upper
bound of the lightest resonances in $\pi\pi$ scattering.  A recent
work~\cite{Polyakov:2001ua} has shown that the upper bound of the
masses of the $\rho$ and $\sigma$ mesons can be expressed in terms of
the LECs $L_2$ and $L_3$.  In particular, the following expression for
the upper bound of the $\sigma$-meson mass was derived:
\begin{equation}
M_{\sigma} < 665[1+0.44 \Delta +0.33 \Delta ^{2} 
+ {\cal O}(\Delta ^{3})] {\rm MeV},
\label{Eq:Delta}
\end{equation}
where
\begin{equation}
\Delta = - \frac{2L_2 +L_3}{L_2}.
\end{equation}
In fact, the ratio $\Delta$ is determined by the $\pi\pi$ scattering
length as follows~\cite{Nagels:xh}: 
\begin{equation}
\Delta = -3 \frac{a^{2}_{2}}{a^{0}_{2}} + {\cal O}(m^{2}_{\pi})
\approx -0.2 \pm 0.6,
\end{equation}
where $a^{2}_{2}$, $a^{0}_{2}$ denote the $D$-wave scattering length
for $I=0$ and $I=2$, respectively.  Though it is hard to judge models
based on this empirical value because of the large error, it is still
of great interest to see the difference between models.  While the
present results are similar to those obtained from other 
models, the ratio $\Delta$ in Eq.(\ref{Eq:Delta}), which is an
important quantity to determine the upper limit of the resonances, 
distinguishes the models.  In Fig. 5, the dependence of the ratio
$\Delta$ on $M_0$ is drawn.  While the result with the form factor in
Eq.(\ref{Eq:types}) shows relatively mild dependence on $M_0$, those
with the dipole and Gaussian form factors depend strongly on $M_0$.
It can be easily understood from the dependence of the $L_2$ and $L_3$
on $M_0$ as drawn in Figs. 3 and 4. 
\begin{figure}
\begin{center}
\includegraphics[height=7cm]{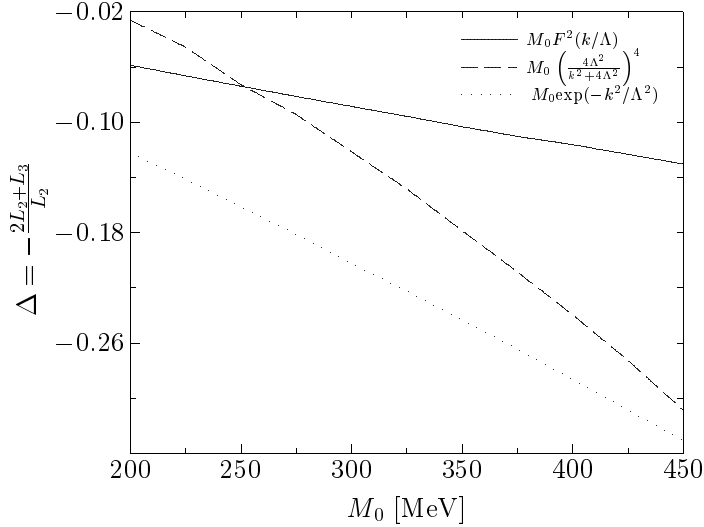}
\caption{The dependence of $\Delta$ on $M_0$.  The solid curve stands for
the result with the form factor from the instanton vacuum
given in Eq.(\ref{Eq:types}),
the dashed one draws the result with the dipole type $M(k)$,
and the dotted one designates the result with the Gaussian one.}
\end{center}
\end{figure}

In Table II we list the results for $\Delta$ and the upper limit of
the sigma meson mass. 
\begin{table}
\begin{center}
\caption{$\Delta$ and the upper limit of $M_\sigma$.}
\begin{tabular}{cccc} \\ \hline
& $2L_2 +L_3 (\times 10^{-3})$ & $\Delta$ & $\le M_\sigma$(MeV) \\
\hline \hline 
local $\chi$QM & 0 & 0 & 665  \\ \hline
Type.1    & 1.67  & -0.103  & 637.2 \\ 
Type.2    & 0.29  & -0.178 & 619.9  \\  
Type.3    & 0.387 & -0.243 & 606.9  \\ \hline
Arriola   & -1.31 & 0.672 & 960.7 \\
VMD(Ref.\cite{Ecker:yg})       & -1.1  & 0.5 & 866.2 \\ 
Holdom(1) & -0.3  & 0.154 & 715.3 \\
Holdom(2) & -0.3  & 0.167 & 720. \\
Bolokhov et al. & 0 & 0 & 665 \\
Alfaro et al. & 0 & 0 & 665 \\
\hline
\end{tabular} 
\end{center}
\end{table}  
As shown in Table II, we can find a very interesting fact: Except for
the present model, all other models presented here give negative
values of $\Delta$.  As a result, while the present work gives the
upper limit of $M_\sigma$ below 640 MeV, all other models in Table II  
predict it rather large.  In particular, Ref.~\cite{RuizArriola:gc}
gives a fairly large value of the upper limit of $M_\sigma$; 961 MeV.      
Though the models of Ref.~{Holdom:iq} contain the momentum-dependent
quark mass, their values of $\Delta$ are quite different from the
present one.  Thus, the values of $\Delta$ distinguish the present
work from other models.        

\section{Conclusions}
In the present work, we investigated the ${\cal O}(p^4)$ effective
chiral Lagrangian in the chiral limit, based on the nonlocal
chiral quark model derived from the instanton vacuum.  Starting from
the effective chiral action, we carried out a derivative expansion
with respect to the pion momenta in order to get the effective chiral
Lagrangian to order ${\cal O}(p^4)$.  The low-energy constants (LECs)
which encode QCD dynamics have been obtained.  We calculated the LECs,
employing three different types of $M(k)$.  The LECs are insensitive
to the types of the form factors.  We found that the results  
are in a good agreement with the empirical data.  Though they are not
much different from those of other models, the present results for the
ratio $\Delta$ turn out to be rather different from them.   
   
A full investigation into the low-energy constants including SU(3)
symmetry breaking and external fields is under way.  
\section*{Acknowledgments}
HCK is grateful to K. Goeke, M. Polyakov, P. Pobylitsa, and
M. Musakhanov for valuable discussions and critical comments.  The
present work is supported by the Korean Research Foundation
(KRF\--2002\--041\--C00067).

\end{document}